\documentclass[10pt]{article}

\title{Microeconomic co-evolution model for  financial  technical analysis signals}
\author{G. Rotundo$^{a}$\footnote{Corresponding author. Faculty of Economics, University of
Tuscia, via del Paradiso 47, 01100 Viterbo, Italy. Tel.:+39 06
4976 6277 Fax:+39 06 4976 6765.  \newline {\em E-mail addresses:}
giulia.rotundo@uniroma1.it (G. Rotundo) marcel.ausloos@ulg.ac.be
(M. Ausloos).},   M. Ausloos $^b$ \\ $^a$ Faculty of Economics,
University of Tuscia, Viterbo, Italy \\ $^b$ SUPRATECS, B5,
Sart-Tilman, B-4000 Liege, Euroland  }

\usepackage{graphicx}

\begin{document}
\maketitle

\begin{abstract}
     Technical analysis (TA) has
been used for a long time before the availability of more
sophisticated instruments for financial forecasting in order to
suggest decisions on the basis of the occurrence of data patterns.
Many mathematical and statistical tools for quantitative analysis
of financial markets have experienced a fast and wide growth and
have the power for overcoming classical technical analysis
methods. This paper aims to give a measure of the reliability of
some information used in TA by exploring the probability of their
occurrence within a particular $microeconomic$ agent based model
of markets, i.e.,  the co-evolution Bak-Sneppen model originally
invented for describing species population evolutions. After
having proved the practical interest of such a model in describing
financial index so called avalanches, in the prebursting bubble
time rise, the attention focuses on the occurrence of trend line
detection crossing of meaningful barriers, those that give rise to
some usual technical analysis strategies. The case of the NASDAQ
crash of April 2000 serves as an illustration.
\end{abstract}

\noindent{\em Key words} SOC model, Technical analysis, large
financial crashes. \newline {\it PACS} { 01.75.+m Science and
society - 89.90.+n Other areas of general interest to physicist }

\section{Introduction}
    Quantitative analysis of financial
market data has well assessed several properties like the  long
term memory in volatility \cite{bm,bcdl,cf,lo}, returns
\cite{deg,dg,dg1}, speculative bubbles \cite{sh}, and the presence
of fractals \cite{av} that has been extensively studied since a
pioneering paper \cite{MN}. Many mathematical and statistical
models \cite{granger,gra1,krawiecki,lz,zp,za} are available
nowadays for a phenomenological description of financial data,
while rigorous theoretical frameworks have  shown to be able to
encapsulate some conjectures like the Elliot waves \cite{EW}.

Alongside the descriptive analysis of macroeconomic quantities,
    theories derived for complex systems can explain the
aggregate behavior of markets through the analysis of its
components at the microeconomic level. Microeconomic models of
financial markets rank in complexity from the simplest models,
typically considering the interaction of two main types of agents
- the fundamentalists and the chartists \cite{cr,ki,k1,kt}- to the
most heterogeneous types of agents;
 an intermediate step
considering the presence of noise traders that act either without
market information or  not caring about the fundamentals, thus
creating white noise, while mean reversion effects \cite{MRP} can
be accounted due to the activity of fundamentalists. The first
question to be raised is whether a microeconomic approach can be
found based on insight about the mechanism of the formation of
financial quantities. If investigations of micro- or
macro-economic models rely on simulation frameworks whenever more
theoretical tools are not available, the evaluation of investment
strategies driven by models, even empirical ones, like those
leading to technical analysis (TA) is a need. Nevertheless it is
difficult to implement them, even through simulations of multi
agent systems, because of the lack of reliability of the
parameters. Indeed it is not easy to perform computer simulations
of markets with interacting agents that trigger their orders on
the basis of technical analysis patterns because technical
analysis rules are more complex than those commonly assigned to
chartists and fundamentalists in computer simulations. Moreover,
to get the best trading decision is still an art, independently
from a model sophistication; indeed the interpretation of charts
heavily relies on the expertise of the analyst.

Therefore we study model property and statistics instead of trying
to draw results relying on heavy computer simulations of a multi
agent system.

 On the other hand, a decision based on financial signal
technical analysis  must take into account the  temporary
occurrence of several patterns. However to start the study of the
occurrence and of the reliability of the simplest components is a
compulsory step towards the comprehension of more complex
configurations. This can  in turn lead to a systematic assessment
of the expertise of such a kind of market analysts.

A cornerstone for technical analysis comes from the expertise of
Charles H. Dow that developed the set of methods that are gathered
under the name of Dow theory.
    Dow theory  \cite{Mu}  considers major trends as those lasting
     more
than one year. Intermediate trends are those that range from a
minimum of three weeks to a maximum of several months, as those
which
  can be useful in futures markets. Short trends can be
identified for time intervals
    shorter than two or
three weeks. Thus it is very important to decide upon a reliable
time interval for implementing a strategy, before trying to define
any trend. Statistics of trend lines will be exploited on the
aggregate of the proposed microeconomic model and compared with
the results obtained on raw data. Such analyzes  should show their
power at their best when performed during periods of high risk
exposure. Among them the rising part of speculative bubbles of
market indices, due to endogenous causes, has been chosen here
below because of the availability of already well assessed
theories \cite{ausloosnikkei,jls3,js3,abmv,macrashPT}. It is worth
remarking that stock market indices actually are a
 weighted mean of stock prices. To perform buy/sell strategies on
 stock market indices (eventually triggered by TA signals) has the
 meaning to buy/sell a previous selected financial product replica of
 the index (Exchange Traded Funds (ETF), certificates).

 Therefore the aim
of this paper is twofold. The first task  is to set up a
microeconomic approach based on insight about the mechanism of the
formation of financial quantities; the second target is to show
how to use the property of the aggregate rising from the model
structure in order to evaluate  the reliability of already often
used methods like those found by chartists in so called technical
analysis (TA) \cite{TA}. In particular the analysis will focus on
the probability estimate of the occurrence of trend lines slopes
and on the estimate the probability of trend lines crossing.

The paper is organized as follows. The next section shortly gives
an overview of the main properties of the NASDAQ July 2000 crash,
of its statistical properties, and shows the bases of the models
that we are going to apply and how to combine them for data
modeling. Sec. 3 introduces TA signals of interest, in particular
so called barriers. Sec. 4 shows how to use the model information
in order to set up a tool in order to estimate both the occurrence
of barrier crossing and the formation of a trend line.

Sec. 5 serves as a conclusion and suggestions for going beyond the
present work. It will appear that the numerical values used to
build the agent-based model describing the financial index are
those of the $2$-dimensional square lattice Bak-Sneppen
coevolution model  \cite{bak}. For completeness the
$1$-dimensional case is treated in an Appendix.

\section{Microeconomic model}

This section aims to set up a model for the rising part of
speculative bubbles due to endogenous causes in order to capture
data features   as the property of long term memory, the
distribution of the size of fluctuations around at the mean, and
the main trend. The modelization tasks can be accomplished through
several models,  depending on the properties of the time series
that need to be maintained. Already existing models for
speculative bubbles \cite{ausloosnikkei,jls3,js3,abmv,macrashPT}
do provide a deterministic function for the main trend and
oscillation modeling through log-periodic patters, but they don't
capture some residual correlation.  Let them be improved as below.
\newline

\subsection{Large financial crashes}
  The theory of
speculative bubbles due to endogenous causes has been extensively
examined indicating self similarity in market indices. Widely
developed numerical studies have extracted common features of
bubbles providing a range for the most important parameters,
taxonomy of bubbles and investigation about signatures for bubbles
due to endogenous causes \cite{js}. Anytime the amplitude of the
crash is proportional to the total price then there is a strong
indication for modeling the logarithm of the index value
\cite{js3,jsor,sz}, instead of using the index itself
\cite{js,Fama}.

 The
modelization of large financial crashes as critical points
\cite{jls3} and a subsequent simplification driven by universality
assumptions \cite{abmv,macrashPT} lead to the approximation of the
main trend of the logarithm of the stock market index w.r.t. the
time to crash $t_c-t$ given by
\begin{equation} \label{Iorder}
F(t)= A+B\ln(t_c-t)\end{equation}
 where $t_c$ is the most probable crash
time and $A$, $B$ are parameters to be estimated via numerical
optimization.

In order to show an example and for further reference below let us
sum up the (speculative) bubble of the NASDAQ that collapsed into
the crash of April 2000 (Fig. 1).

Accordingly to \cite{js3,js} let the financial signal data
$\{y(t)\}_{t=1,T}$ be the logarithm transformation of the NASDAQ
100 Composite  index daily closing value between Jan. 01, 1997 and
March 10, 2000, i.e. $T=833$.
  The best fit of (\ref{Iorder}) to the ascending part of the bubble has
been performed using the minimum least squares method. The results
are $A=7.91$, $B= -0.54$, and $t_c$ corresponding to July
4$^{th}$, 2000 \cite{js,ver200}. Notice that the actual crash
date, April $11^{th}$ approximately occurs three months before the
$t_c$ estimated by the fit. We have noticed on other data as well
that this is usually experienced when (\ref{Iorder}) is used.

 Evolution models more complex than (\ref{Iorder}) can be considered for the description of the
main oscillations \cite{js}, but the scaling of correlations
changes only slightly. The periods that correspond to the  rise
and to the successive burst of a speculative bubble due to
endogenous causes are characterized by several oscillations around
the main trend \cite{ausloosnikkei,drodz,feigenbaum,kaizoji}, as
widely examined through the papers that assess the similarities of
large financial crashes properties with earthquake phenomena
\cite{js} or sandpile avalanches on fractal structures
 \cite{ausloosnikkei,vdwdhumaPRE}. Although the importance of log periodic accelerating
oscillations going close to the most probable crash time is deeply
connected with the self similarity hypothesis, and discrete scale
invariance, its validity is still debated, because the residual
correlation evidences the role of residual noise, pointing to the
limit of the theory and suggesting to look for other models for
describing the major fluctuations.

 Let the so called residuals $R(t)$ be defined through (see Fig.
1(b) for a display)

\begin{equation} \label{Iorder_res} R(t)=\exp(y(t))-\exp(F(t)).
\end{equation}

It is usually found that most market index data $R(t)$ show long
term memory correlation indicating a mean reverting process
\cite{MRP}. This implies that a useful statistical property to
compare models to real data should be found in the Hurst exponent.
The Detrended Fluctuation Analysis (DFA) technique
\cite{maklo,DFA2,DFA1} is often used in order to characterize
fluctuation correlations in such time series, through the power
law exponent $\alpha$. In this NASDAQ case study
$\{R(t)\}_{t=1,T}$ is characterized by $\alpha=1.39
(1.37,1.40)$\footnote{For each variable empirically estimated the
numbers inside the parentheses are the $95\%$
confidence interval.} 
corresponding to a Hurst exponent $H \simeq 0.39 (0.37,0.40)$.
Thus the residuals $R(t)$ can be modeled through a theoretical
fractional Brownian motion (fBm). In this case the time at which
the random walker starting at the origin first returns to the
origin, i.e. be the first return time $T$ of a fBm   has the
following probability decay \cite{dy,rdfirst}:
\begin{equation} \label{PT} P(T) \sim T^{H-2}.\end{equation}
The estimate of $P(T)$ on $\{R(t)\}_{t=1,T}$ data at the level of
the initial time value $R(1)$ in the NASDAQ case gives $
H=0.42(0.06,0.77)$ and fine agreement with the above estimated $H$
through the DFA (see Fig. 1(c)).

\subsection{The Bak and Sneppen model}
The simple Self-Organized Criticality model of  Bak and Sneppen
(BS)
    \cite{bak,rj,sk} has been shown to fulfill not only characteristics of species
    distribution evolutions like a power
law in the distribution of avalanches, but also the
    characteristics required for earthquake modeling, both for spatial and
temporal correlation functions,  or also in landslides
\cite{guz1,ik,lm}. It is of great interest that the probability of
the occurrence of the next avalanche can be estimated, as it
emerges from the model dynamics. Interestingly earthquake models
were shown to be well suited for the description of market data
that are characterized by cascade crashes and are followed by slow
recoveries \cite{kr,abmv}. This behavior has been detected in
several speculative bubbles and attributed to endogenous causes
\cite{sh}. At each time $t$ the $d$-dimensional BS model deals
with $L^d$ species that compete for their survival.

In a financial application of the BS model each species can
represent either an agent or a class of them through a
representative agent. At time $t$ each species is fully described
by its fitness $f^d_i(t)$, $i=1,\cdots, L^d$ drawn at time $0$
from a uniform distribution in $[0,1]^d$. A change in fitness of
one species implies an evolution of others as well: there is
co-evolution. In the original evolution model the fitness can
represent either the living capability of the species or the
population, or a barrier to be overcome. The average fitness can
be defined as the simple mean \cite{lc1,lc2,lc3} of the individual
fitness, i.e.,
\[\bar{f}^d(t)=\frac{1}{L^d} \sum_{i=1}^{L^d} f^d_i(t) .\]
The BS model has been introduced for simulating the collective
behavior of interacting groups or individuals. In financial
markets each $f_i$ can be interpreted as the estimate of the
market price by either groups or agents.
On financial markets this mean fitness $\bar{f}^d(t) \in [0, 1]$
can be used as an approximation of the market index value,
resulting from the market price, as seen by/at each agent level,
scaled to the $[0, 1]$ interval as it is built up through the
components of the market and the agent behaviors. Thus the $L^d$
species can be interpreted as $L^d$ groups of investors, more
properly than just $L^d$ single investors, whose contribution to
the formation of the price at time $t$ is given by $f^d_i(t)$ that
can include the raw price $p^d_i(t)$ (normalized to $1$) as well
as the raw price already multiplied by the weight of the group due
to its social impact $f^d_i(t)=\omega^d_i(t)\,p^d_i(t)$, $\sum_i
\omega^d_i(t)=1$.

Groups (agents in the smallest size case) are thus modeled as
organized in a simple social lattice or network that in dimension
$d$ connects each group only to his first $2d$ nearest neighbors.
The usual boundary conventions of the Bak-Sneppen model hold. We
are aware that this model should be considered as a high
simplification of a social/financial network that keeps trace only
of the
 most important influences of a group over the other.
Extensions to more complex networks imply further work depending
on possibly unrealistic features to be found in our
$d$-dimensional model.

Let at each time step  the group with the lowest
price\footnote{It could be the value the furthest away from the
market price at time $t-1$, see
 \cite{ACPBS} for such an alternative in macroeconophysics} randomly "adapt", i.e.,
  change the price and
affect its nearest neighbors in the spirit of the BS model.
Extremal values are those the furthest away  from the mean. The replacement
of the lowest value $f_i$ by a random number can be interpreted
  as the correction to the worst price underestimate.
When
$L \rightarrow \infty$, and for $t$ large enough  almost all
species have their fitness above a threshold $f^d_c$
\cite{pmb,yam}; these fitnesses are therefore uniformly
distributed in $(f^d_c,1)$.

   An evaluation of the ''distance'' of this simple toy model
hypotheses and implications  from  true social/financial systems
with complex interactions and imitative behaviors is nearly
impossible. Nevertheless it  is interesting to note that the
threshold for $\bar{f}^d(t)$ becomes $\bar{f}^1_{c}=0.83351$ if $d=1$   
and $\bar{f}^2_{c}=0.66443$ if $d=2$ \cite{lzg} (Fig. 2(a)), as
noticed by Li and Cai \cite{lc2}, reasonably similarly to what is
found in some social systems.
 An interesting
behavioral remark is that the critical threshold in the case $d=2$
fits social rules that assign a special weight to decisions when
approximately $2/3$ of people agree.

\subsection{Avalanches: degradation and recovery}
At this stage it is of interest to stress that the BS model has
led to several definitions of avalanches \cite{bak,lc2,pmb}. The
duration of an avalanche in the original BS model refers to the
time spent by the lowest $f_i(t)$ below $f_c$. On the other hand,
Li and Cai define an avalanche (duration) as the time spent by
$\bar{f}^d(t)$ below $\bar{f}^d_c$. This is in fact only
degradation part of the whole BS avalanche. This definition
neglects part of the signal, i.e. the time spent by the signal
$\bar{f}^d(t)$ $above$ the threshold. This signal is of interest
as well in particular in financial and social matters. One could
define and analyze the statistics  of time intervals between
maxima (in/and minima) of the signal. These would encompass cycle
like situations containing degradation $and$ recovery processes.
In the present paper the Li-Cai avalanche definition will be used,
leaving other definition investigation for other work.

 An important feature concerns the structure of such avalanches.
After the first transient phase the model dynamic leads to the
activity of the system characterized by $\bar{f}^d(t)
>\bar{f}^d_{c}$
 \cite{lzg}.  Following the definition reported in
\cite{lc1,lc2}, the size $s$ of the degradation part of avalanches
is defined as its temporal duration, i.e. an avalanche of size $s$
remains below $\bar{f}^d_c$ for $s-1$ time steps $\bar{f}^d(t)$.
Thus the duration is the number $s$ such that
\[ \bar{f}^d(t) > \bar{f}^d_{c}, \bar{f}^d(t+1) < \bar{f}^d_{c}, \cdots, \bar{f}^d(t+s-1)
< \bar{f}^d_{c}, \bar{f}^d(t+s) >\bar{f}^d_{c}.\]

 It has been shown \cite{lzg,sjb}
that  the $s$-distribution of such degradation part of avalanches
follows a power law
\begin{equation}\label{scala} P(s) \propto s^{-\tau}.
\end{equation}

For the $1$-dimensional BS model $\tau=1.8$; for the
$2$-dimensional square lattice BS model $\tau=1.72$ \cite{lc1}.
Moreover although the average avalanche size depends on the
distance from $\bar{f}^d_c$,  the values of $\tau$ are independent
of the level chosen \cite{lc2} for an infinite system.

For the purposes of TA, and trend line property searches we also
studied the recovery situation, identified by the permanence of
the signal above the threshold $\bar{f}^d(t)$ in the sense of
Li-Cai. In order to do so a mirror-like situation must be
envisaged.

 Of course $-\bar{f}^d(t) $ has
the same long memory degree as $\bar{f}^d(t)$; moreover it is
possible to define the recovery size $s$ as being a sequence of
$s$ time steps such that
\[ -\bar{f}^d(t)< -\bar{f}^d_{c}, -\bar{f}^d(t+1)> - \bar{f}^d_{c},
\cdots, -\bar{f}^d(t+s-1) > -\bar{f}^d_{c}, -\bar{f}^d(t+s) <
-\bar{f}^d_{c}.\] Because of the fact that\[P(  -\bar{f}^d(t)<
-\bar{f}^d_{c}, -\bar{f}^d(t+1)> - \bar{f}^d_{c}, \cdots,
-\bar{f}^d(t+s-1) > -\bar{f}^d_{c}, -\bar{f}^d(t+s) <
-\bar{f}^d_{c} )\]\[= P(\bar{f}^d(t) > \bar{f}^d_{c},
\bar{f}^d(t+1) < \bar{f}^d_{c}, \cdots, \bar{f}^d(t+s-1) <
\bar{f}^d_{c}, \bar{f}^d(t+s) >\bar{f}^d_{c} )\] the scaling of
the recovery time span maintains the property (\ref{scala}).

\subsection{The BS model applied to residuals of  financial indices}

As seen in the previous section the BS model provides a
description for the distribution of species evolution avalanches.
It is of interest to consider them as the analog of  those that
are found before a large financial crash and are compatible with
the recoveries to the mean trend observed on usual data; thus the
BS model provides an interesting modelization for the oscillations
of the residuals. Of course the range of $\bar{f}^d(t)$ must be
properly rescaled in order to fit the range of $R(t)$.

The self similarity degree of the simulations obtained through the
DFA on the $1$-dimensional BS model gives a self similarity
exponent $H=0.07$, quite far from the value of $R(t)$ of the
NASDAQ case study, whilst the same analysis performed on the
stable phase of the $2$-dimensional BS model gives values
approximately normally distributed with mean $\bar{H}=0.277$ and
standard deviation $\sigma_H=0.088$ (Fig. 2(b)). Taking into
account that numerical estimates are biased by errors due to
finite-size of the sampling, as it emerges also for $\tau$ (Fig. 2
(c)), the above results allow to state that the trajectories
$\bar{f}^d(t)$ obtained through the $2$-dimensional square lattice
BS model can replicate the self similarity degree of case studies
like the NASDAQ residuals, and constitutes a better choice than
the $1$-dimensional BS model.

A further empirical analysis looking for the subsequence of NASDAQ
data that best fits the $2$-dimensional BS model parameters $H$
and $\tau$ has been performed. The residuals time series starting
since December 23$^{th}$, 1998 (see the vertical line in Figs.
1(a) and 1(b)) till the end shows parameters
 and
$H = 0.28(0.25, 0.31)$
$\tau=1.48(1.22, 1.74)$ (Fig. 1(d)) and it is the part of the NASDAQ data that
 best fit both the $H$ and $\tau$ BS model
parameters  (see Fig. 1(a)).

 Hereinafter let $\{\bar{f}^d(t)\}_{t=1,T}$ be a sampling
from the $2$-dimensional square lattice BS model in the stable
phase.
\newline The $1$-dimensional case is briefly worked out in
Appendix A. Thereafter we drop the index $d=2$ for simplicity in
the writing.

 The detection of a mean reverting process
allows us to look for the parameters $\theta$ and $\gamma$ such
that
\begin{equation}\label{barfz}
 \theta + \gamma \bar{f}(t) \end{equation}
has the same self-similarity degree $H$ and the same range as
$R(t)$, thus explaining the oscillations as due to the most
important social interaction links of each agent. Recall that
practically
\begin{equation} \label{g}
\gamma=\frac{\max (R(t))- \min
(R(t))}{\max(\bar{f}(t))-\min(\bar{f}(t))}
\end{equation}
and
\begin{equation} \label{t}
\theta=-\gamma  \min(\bar{f}(t))+\min (R(t)).
\end{equation}

 Recalling (\ref{Iorder_res}) we have that
\begin{equation}\label{rec1}
g_1(t)=\exp(F(t))+\theta+\gamma \bar{f}(t)\end{equation}
constitutes a model for market indices during the rise of
speculative bubbles that replicates  the deterministic exponential
trend, the avalanche exponent $\tau$ and the $H$ self-similarity
exponent of the NASDAQ index. The faster than exponential growth
is typical of speculative bubbles due to endogenous causes.

The modelization of  the NASDAQ $R(t)$ through a fBm leads to the
first return time  probability decay exponent equal to $H-2$. Thus
(\ref{PT})  measures the decay of the size of periods passed
either over or under the value of the process at the initial time,
henceforth including, but not being limited to, the avalanches as
defined in the BS model within the Li-Cai description. However the
agreement of the exponent $\tau=1.72$ of the $2$-dimensional
square lattice BS model and the exponent $2-H$ into the  $95\%$
confidence interval in the case of the NASDAQ provide a further
validation of the choice of the $2$-dimensional  square lattice BS
model.

The function
\begin{equation}\label{rec2}
g_2(t)=\exp(F(t))+\theta'-\gamma' \bar{f}_c\end{equation} is also
well suitable for our modelization proposal, imposing that

\begin{equation}\label{barfz1}
 \theta' - \gamma' \bar{f}(t) \end{equation}
has the same self-similarity degree H and the same range as
$R(t)$. Parameters $\gamma'$ and $\theta'$ are calculated by using
formula (\ref{g}) and (\ref{t}) where $\bar{f}(t)$ was substituted
by $-\bar{f}(t)$.

The recovery time scale distribution to the
function obtained substituting $\bar{f}(t)$ by $\bar{f}_c$ is
equal to the avalanche time scale distribution calculated for
$g_1(t)$ under the same substitution.
The best fit to the data starting on December 23$^{th}$, 1998
leads to the same parameters $\gamma$, $\gamma'$, $\theta$,
$\theta'$ because the maximum and the minimum of $R(t)$ occur
after December 23$^{th}$, 1998 (see formula (\ref{g}) and (\ref{t})).

Table \ref{tab:tau} resumes the values of NASDAQ avalanches under
$g^{+}=\exp(F(t))+\theta+\gamma \bar{f}_c$, and of recoveries of
$g_2(t)$ to $g^{-}=\exp(F(t))+\theta'-\gamma' \bar{f}_c$, that
correspond to the critical level $\bar{f}_c$ for $\bar{f}(t)$.
 Figs. 3(a) and 3(b)   show samplings of (\ref{rec1}) and (\ref{rec2}).
Notice the mirror symmetry  w.r.t.  $g^{+}$ and $g^{-}$.
We are going to use $g_1(t)$ in order to model avalanches, and
$g_2(t)$ in order to model recoveries. We are going to use  the
above results in Sec. 4 in order to give an estimate of the
probability of $g^{+}$ and $g^{-}$ crossing and trend line slope
and location.

\section{Technical analysis signals}

     Technical analysis \cite{TA}  is based on     the reaction of
financial agents to market conditions as they can be detected
through the study of charts, i.e. financial market data plot. It is characterized by the  usage of
particular signals in order to trigger buy/sell orders.

  The most significant
criticism against such a technique is its possible lack of
precision in the recognition of signal patterns and its subjective
judgement in their interpretation. These could mislead to the
precise timing of  signals. In spite of this uncertainty it is
worth remarking  that the methods have been surviving and
developing for  a long time. This consideration suggests to look
for the extraction of the essence of the methods not affected by
any psychological effect but which can be detected by an automatic
decision support system properly calibrated.

TA signals are triggered by the occurrence of some patterns, that
can be separated further on.
  The easiest figures to deal with and which can provide useful trading
  information are horizontal barriers. An immediate further step is given by the
crossing of lines with a not null slope, like trend lines, fan
lines, and channels. Trend lines are straight lines joining  sequences of at least two
 minima (maxima) with the second one higher (lower) than the first
 one; fan lines are trend lines joining two points: the first one is kept fixed (it is common
 to all of  them), and it is a minimum (maximum), while the second point is given by the
 subsequent minimum (maximum). Channels can be drawn in the case for which the data
 exhibits a  sequence of minima with linearly growing height and a sequence
 of maxima with approximately the same linearly  growing height. In this case the line
 fitting the sequence of maxima and the line fitting the sequence of
 minima identify a channel. The identification of the above quantities is
highly sensitive to the time scale that is chosen, as moving
averages and their relative crossing are \cite{mamav}.

   Buy/sell signals rely on the identification of particular
configurations. Several rules are known for their identification
\cite{Mu}. They   provide a set of buy/sell triggering orders more
complex than the simple crossing of an horizontal barrier. However
the knowledge and the understanding of the base components are the
starting points towards the analysis of more complex patterns.

Recently TA has been reconsidered and strategies redefined in
order to take into account not only the price variation but also
the effect of volume
    bearing upon classical mechanics ideas
    \cite{mavolume1,mavolume2}. Although the volumes play an important role
    in technical analysis their examination is also outside the scope of
    this paper.

\section{Self-barrier crossing}
As recalled here above,  the probability  of avalanche duration in
Self-Organizing Critical systems can be used for modeling the
probability of falls in markets, thus carrying on the comparison
with the earthquake theory, as it has been evidenced across the
literature on large financial crashes \cite{macrashPT}.

Here below, it is shown how to use the property of the model
previously discussed in order to estimate the probability of the
expected time for line crossings.
 For this crossing search we use the data self generated values.
 i.e. $g^{+}$ and $g^{-}$ defined in Sec. 2.4 so that we call the problem self-barrier
 crossing in analogy with self-avoiding walk wording \cite{saw}.

\subsection{Line crossing estimate based on the model structure }

Let $l(t)=a\,t+b$ be a straight line, see Fig. 4. Let $g(t)$ be
either $g_1(t)$ or $g_2(t)$. The average
\[q(t)=<g(t)>\] is equal to \[q(t)= \exp(F(t))+\theta+\gamma <\bar{f}(t)>\]
in the case $g(t)=g_1(t)$, and to
\[q(t)=\exp(F(t))+\theta'+\gamma'<-\bar{f}(t)>\] in the case
$g(t)=g_2(t)$. The function $q(t)$ is a deterministic function of
$t$, and the expected intersection time $t=t^*$ for the line
crossing can be calculated by
\begin{equation} \label{ql}q(t*)=l(t*).\end{equation}

In the case $g(t)=g_1(t)$ the error on $t^*$ can be deduced from
$\gamma Var( \bar{f}(t))$. This provides an estimate about the
spread around the expected time for the process to cross any line
(Fig. 4).

Since it is known that the maximal change in $s$ steps, $\forall t
\in [1,T]$,
 is $\max \mid \bar{f}^d(t+s)-\bar{f}^d(t)\mid5s/L^d $ \cite{lc2} it is found that  the $\max$ and $\min$ slope
between two points  $(t,g_1(t))$ and $(t+s,g_1(t+s))$ are given by
\begin{equation} \label{slope}( e^{F(t+s)}-e^{F(t)}\pm \gamma
\frac{5s}{L^d}) \frac{1}{s}. \end{equation} Analogous results hold
for $g(t)=g_2(t)$.  In order to tighten bounds on slopes let us
analyze the distribution of the slopes of lines joining
$(t,\exp(y(t)))$ and $(t+s, \exp(y(t+s)))$, $(t, g_1(t))$ and
$(t+s, g_1(t+s))$, and $(t, g_2(t))$ and $(t+s, g_2(t+s))$,
respectively, for $s=2, 3, 4, 11, 13$, that are values significant
for the NASDAQ avalanches and recoveries.

The analysis is carried on both   the entire NASDAQ time series  (Figs. 5(a), 6(a), and
7(a)) and  the best fitted subperiod starting on December
23$^{th}$, 1998  (Figs. 5(b), 6(b), and
7(b)).
 Due to the avalanche
distribution the most frequent value is lower than the median,
that is lower than the mean. Fig. 5 reports the histograms and
their cumulative distribution. For each fixed time step $s$ the
frequency of the slope of lines joining points with time distance
$s$ (either on the raw data or on the simulated ones) can be
obtained directly from the histogram. A trader would examine how
many percentage of the slopes is between two bounds in order to
estimate the risk of some strategy.

As an example, referring to Fig. 5(a), the $75\%$ of the slopes are
in a interval of width the standard deviation around at the mean,
whilst $10\%$ are in a more tight
 interval with width $2.3$  (see Fig. 8).

On the entire time series both the Lilliefors  and the Jarque-Bera
test reject the normal hypothesis distribution at $95\%$
confidence level in all the NASDAQ cases and for the biggest
values of $s=11$ and $s=13$ also on $g_1$ and $g_2$ (Fig. 6(a)).
On the NASDAQ data set starting on December 23$^{th}$, 1998  both
tests reject the normality hypothesis only for $s=2$ and $s=13$
(Fig. 6(b)). Table 2 resumes the values of mean and standard
deviation on both the entire time series and on data since
December 23$^{th}$, 1998. It is worth noting the relationship
between the slopes distributions of the NASDAQ, $g_1$, and $g_2$
(Fig. 7).
  Their comparison  shows the
probability of the occurrence of slopes joining points at time
step distance corresponding to the most frequent avalanche size,
to the median, and to the mean. The NASDAQ distribution is higher
around at its mean in all the reported samples.
  The frequency of the slopes of $g_1$ and $g_2$ around at
the mean is a lower bound for the NASDAQ slopes. The evaluation of
bounds relevant for particular strategies is left for future work.

In TA trend lines must be drawn by joining either decreasing
sequences of maxima (at least two), or increasing sequences of
minima \cite{Mu}.
 No method seems available up
to now in order to forecast whether the second maximum (minimum)
is lower (higher) than the first one, thus giving rise to a trend
line \cite{Mu}. However because the deterministic part of $g(t)$
is strictly monotonous it is possible to calculate some bounds:
the minimum time $s$ such that $g(t+s)>g(t)$, is given by
$\exp(F(t+s))-\exp(F(t))>\gamma 5s/L^d$. The maximum time $s$ such
that the inequality $g(t+s)<g(t)$ can hold is given by
$\exp(F(t+s))-\exp(F(t))<\gamma 5s/L^d$.

\subsection{Trend line detection}

 Trend lines \cite{Mu} play an important role because they serve as a  basis
for classical technical analysis.
 Such lines are drawn joining a sequence of at least two  maxima
 (minima) with the second one lower (higher) than the first one, each one being
selected as global maxima (minima) on time windows \cite{Mu}. The
size of time windows depends on the information that the analyst
is looking for. Thus to trace out trend lines strictly depends on
the time width that is chosen. Major trends are defined by the Dow
theory  as trends during more than one year, although this limit
can be lowered to six months on the future market. Thus in this
case maxima (minima) can be looked for on monthly time windows. At
the opposite time window size choice there are short time trends,
that are shorter than two $-$ three weeks. Intermediate trends
should take into account periods of two $-$ three weeks up to
several months. They are more stable than those observed on short
time intervals \cite{Mu}, and more meaningful than those over long
time intervals, on which the exponential trend is already visible.

Any upward (downward) peak can be considered as a maximum (a
minimum) and can be used for trend line identification. On the
NASDAQ and on the BS simulations the peaks occur any $2-3$ steps,
thus short time trend lines could be drawn and the results on the
previous section could be directly applied.

However the choice of the time window where to look for global
maxima much relies on the feeling of the analyst. We stress that
the BS model can provide a further contribution about the
occurrence of trend lines in  the case in which the investor is
not interested to peaks inside the avalanche, even if they are
local maxima in their time window.

 Information model$-$related furnishes
estimates on the occurrence of two maxima separated by an
avalanche. Their time steps distance is bigger than the time width
of an avalanche. The function $g_1(t)$ takes into account
avalanches following the definition of the Li-Cai model
\cite{lc1}, and the function $g_2(t)$ considers recoveries.

Henceforth the function $g_1(t)$ can provide information about the
time steps distance of values separated by an avalanche, thus
about the minimal distance between maxima separated by an
avalanche. On the other hand the function $g_2(t)$ can provide
information about the time steps distance of values separated by a
recovery, thus about the occurrence of minima. This kind of
approach can be useful to set up automatic trading rules that
trigger orders on the basis of the occurrence of falls in the
market, or in the case of the end of a small bubble and of the
return to the fundamental price.

The BS model shows a distribution of avalanche sizes which decays
as a power law with the size of the avalanches. Thus the most
frequent value is much smaller than the mean (Fig. 2(c)). An
investor can thus practically estimate the probability to have a
short time avalanche instead of a medium size one directly from
frequencies of avalanches sizes. Because of the asymmetry of the
frequency distribution of an avalanche sizes, that is limited by 0
on its left, and  that has power law tail on the right,  the most
frequent values are the smallest ones ($1$ and $2$ time steps
below the threshold, i.e. avalanche sizes $2$ and $3$); notice
that the median of the avalanche size is lower than the mean (Fig.
9).

The above remarks evidence once more that short time trend lines -
even separated by an avalanche $\bar{f}^d(t)$- are the most
frequent.

Thus downward trend lines can be considered for short time
intervals, but not for medium size ones. Anyway in a growing
market downwards trend lines are surely going to be crossed; the
most interesting question is about the upward trend lines that
join sequences of local minima which height is increasing.

The function $g_2(t)$ can be used in order to give estimates of
distances between minima separated by recoveries. Its structure of
recoveries is mirroring the structure of BS ordinary avalanches.
Thus the mean distance between two minima separated by a recovery
is bigger than the avalanche size. Here again the short time
recoveries are the most frequent ones, but in this case the second
minimum could not be higher than the first one. Due to the
exponential term, the median and the mean size recoveries separate
minima with the second one higher than the first one.

\section{Conclusions}
Any automatic trading system can trigger buy/sell orders when some
either upper or lower barrier is crossed. Rules like these, even
so simple, have been  able to cause crashes in markets. Technical
analysis studies concern a more complex structure of signals and
often rely on the sensitivity of the analyst.

We have observed an analogy between statistical properties of a
coevolution model, in particular the avalanche content
description, with the residuals of a financial index signal like
the NASDAQ 100 Composite - residuals obtained from the latter
signal first order approximation obtained by the theory of  large
financial crashes.
\newline In view of the analytical properties of the
superposition of such a microscopic models, we have been able to
derive estimates for avalanche and recovery duration time. We have
pointed out the qualitative features helping technical analysts to
elaborate more refined approached than classical ones. Several
quantitative features are also available.
Other indices could be used for a search on universality
properties. We stress that we have discussed not only degradation,
but also recovery features.

 The suitability of
the mean fitness properties for bubble models as they emerge from
the BS model dynamics does not stop at the level of numerical
analysis of correlations, but  goes further into the dependence
structure of avalanches, making this approach a sound complement
to the fBm approach. Moreover the BS model furnishes an
agent-based model for an explanation of the cooperative behavior
that leads to the earthquake and large financial crashes
phenomena, thus embracing simulations into a theoretical framework
that allows to state the reliability of the results apart from
numerical instabilities.

The set up of a model for the residuals provides a further insight
on the formation of trend lines.
Statistics drawn on the BS model
about the occurrence of trend lines slopes provide market analyst
by bounds of the probability of the occurrence of trend lines on
market index data. Moreover, once trend lines are drawn, the
probability of their crossing can be estimated from the model.
This
is a first step towards other signal technical analysis  and to
the assessment of the usage of technical analysis rules that
supersede the skill of the single analyst.

In addition the 2-dimensional model, necessarily implying the
existence of more than 2 nearest neighbor agents, obviously 4 for
the square lattice data examined here above,  finds some
correspondence in the sand pile model on a fractal basis
simulating financial avalanches before a crash as studied in
references \cite{ausloosnikkei,vdwdhumaPRE} where the periodicity
of the log periodic oscillations indicate that the relevant number
 of agents is between 3 and 4.

For further work, it could be suggested that more complex signals
be examined, together with a deeper analysis of the fractal and
multifractal structure both of the microscopic model and of market
data.

\vskip 12pt

 {\bf Acknowlegements}

MA thanks  EC Project 'Extreme events: Causes and consequences
(E2-C2)', Contract No 12975 (NEST) for some financial support.

\section{Appendix}

In this appendix we outline a few results corresponding to those
on the main text, but in which a $1$-dimensional model rather than
a $2$-dimensional model is used. We recall that
$\bar{f}^1_c=0.8335$; the value of $\tau$ for avalanches in the
Li-Cai spirit is $\tau=1.80$ \cite{lc2} and it is the same
degradations or recoveries.

The estimate of $H$ fluctuates around at 0.07.
 These
values are rather far from the NASDAQ data indicating poor
agreement with a $1$-dimensional model.

The long term memory property of a time series can be estimated
through its components: it is reported in \cite{granger} that the
sum of two independent fractionally integrated processes of order,
respectively, $d$ and $d'$ is $\max\{d,d'\}$. The relationship
$H=d+1/2$ \cite{em} allows to deal with a fBm $z_H(t)$, $H=d+1/2$
in order to fix $\theta$, $\gamma$, $\zeta$, $H$ such that

\begin{equation}\label{barfz2}
 \theta + \gamma \bar{f}(t)+\zeta z_H(t) \end{equation}

has the same self similarity exponent and the same range of
$R(t)$.

Whilst the contribution of $\bar{f}(t)$ is due to the local agent
interactions, the contribution of $z_H(t)$ can be interpreted
either like the contribution  of noise traders, in the case of
uncorrelated signal, or like the presence of fundamentalist
traders in the market whether a mean reverting process occurs
\cite{MRP}. This gives an interesting perspective about the level
of SOC that can be masked by either noise or fundamentalists
traders. However for the purposals of this paper to give an
evaluation tool for the crossing of lines this model would give
more weak results. As an example in (\ref{ql}) the variance about
the expected crossing time would contain also the variance term
due to the presence of fBm. Also the BS structure of avalanches
would be modified, moreover a model of data based uniquely on fBm
would result more simple, and so preferable to (\ref{barfz2}).
Thus the choice of the  $2$-dimensional BS models meets the task
to give simple description  at a microeconomic level more
meaningful of those related to a generic fBm,
 with the maintainance
of properties (self-similarity, avalanches and recoveries) at the
macro level more suitable for data than the $1$-dimensional BS
model.

\newpage

\begin{table}
\centering \caption{Avalanche and recovery size estimates.
According to definition the minimum sized avalanche stays only one
time step below the threshold, giving rise to avalanche size
$s=2$}
\begin{tabular}{l|l|lll}
\hline\noalign{\smallskip}\label{tab:tau} NASDAQ   & $\tau$ &
avalanche size&  & \\
  & &most frequent & median & mean \\
 \hline
 avalanche below  $g^{+}$ & 0.72 ( 0.18,1.25)    &3(28$\%$) & 4 &13  \\
\hline\noalign{\smallskip} recoveries to $g^{-}$&
$1.79(1.33,2.25)$ & 2(38$\%$)&4 & 11\\ \hline
\end{tabular}
\end{table}

\begin{table}
\centering \caption{Mean and standard deviation (between the
parentheses) of the slopes of lines joining $(t, \exp(y(t)))$ and
$(t+s, \exp(y(t+s)))$ (NASDAQ),  $(t, g_1(t))$ and $(t+s,
g_1(t+s))$ ($g_1$), and  $(t, g_2(t))$ and $(t+s, g_2(t+s))$
($g_2$) for $s=2, 3, 4, 11, 13$: (a) Analysis on the entire time
series; (b) Analysis on the time series since December 23$^{th}$,
1998} \vskip 12pt
 (a)
\begin{tabular}{l|l|l|l|l|l}
\hline\noalign{\smallskip}\label{tab:ms}  data     & $s=2$ &$s=3$
& $s=4$           & $s=11$         & $s=13$  \\ \hline
NASDAQ & $4.51(39.1561)$ &  $4.52(27.3603)$ &  $4.45(22.3469)$ &   $4.24(12.0141)$ &  $4.18(10.8556)$     \\ \hline
$g_1$   & $4.95(44.2339)$ &  $4.90(29.1199)$ & $4.86(22.3350)$  & $4.72(  9.5639)$ & $4.69(8.6526)$      \\ \hline
$g_2$   & $4.04(44.3814)$ &  $4.05(29.4014)$ & $4.06(22.8014)$ &  $3.92( 10.6265)$ & $3.87(9.7463)$     \\ \hline
\end{tabular}
\vskip 12pt

(b)
\begin{tabular}{l|l|l|l|l|l}
\hline\noalign{\smallskip}\label{tab:ms1}  data     & $s=2$ &$s=3$
& $s=4$           & $s=11$         & $s=13$  \\ \hline NASDAQ &
 $9.36(55.5312)$  &   $9.29( 38.3107)$ & $9.05( 30.9240)$  &   $8.42( 15.6117)$  & $8.27(13.8854)$ \\
 \hline$g_1$   &  $8.98(46.0860)$  &    $8.89(30.1768)$  &    $8.71(23.2664)$ &  $8.31(9.7973)$ &  $8.23(8.8547)$ \\\hline
  $g_2$   & $9.53(46.1976)$    & $9.50(30.4889)$    & $9.57(23.8624)$    & $9.21(11.2231)$    &
  $9.09(10.2732)$
\end{tabular}
\end{table}




\newpage

\begin{figure}
\includegraphics[height=6cm]{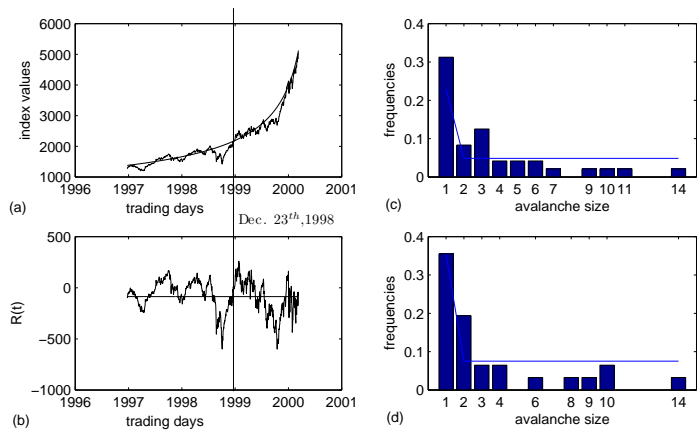}
\caption{The NASDAQ  100 Composite index case study. (a) NASDAQ 100
Composite index daily closing value and its exponential
 approximating function (\ref{Iorder}). The data set used in
order to study the ascending speculative bubble ending at the
crash is for the trading days since January 1$^{st}$, 1997 till
March 10$^{th}$, 2000, for a total   of 833 points. The vertical
line corresponds to December 23$^{th}$, 1998 and emphasizes the
starting day of the period that is best suitable for the BS model.
(b) Plot of the residuals $R(t)$, as from Eq. (\ref{Iorder_res}),
i.e., the difference between the raw NASDAQ 100
 Composite index daily closing value and its exponential approximation. The horizontal line corresponds to the level
 $R(1)$.
 (c) The  estimate of $P(T)$  on the entire time series at level $R(1)$ according to
 (\ref{PT}) gives
$ H=0.42( 0.06, 0.77)$. (d) The  estimate of $P(T)$  on data since
December 23$^{th}$, 1998 till the end gives $ H=0.58( 0.12, 1.04)$
at the level of the first day value.}
\end{figure}

\begin{figure}
\includegraphics[height=8cm]{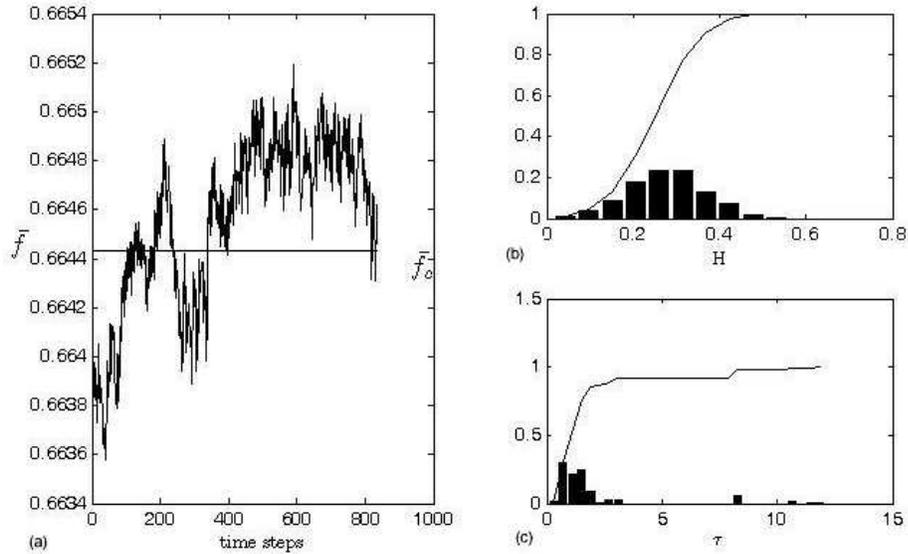}
\caption{ $2$-dimensional BS trajectory in the stable phase: (a) A
sample  and the critical level $\bar{f}_c$. For this sample
$H=0.38 (0.35,0.40)$,
 $\tau=0.05 (-1.89,1.99)$. (b)
 Estimate of the self similarity degree $H$ through the
DFA analysis on $1000$ trajectories of the $2$-dimensional BS
model during the stable phase and its cumulative function. The
mean is $0.277$ and the standard deviation is $\sigma_H=0.088$.
The frequency of BS trajectories that have the $H$ exponent
confidence interval overlapping the  confidence interval of the
 NASDAQ $H$ on the entire time series is approximately $17\% $, while on the period since Dec. 23$^{th}$ it is approximately $24\%
 $.
 (c) Estimate of the avalanche exponent $\tau$  on
subsequences with length $T$ of trajectories of the
$2$-dimensional BS model during the stable phase and its
cumulative function. The mean is $1.9$ and the standard deviation
is $\sigma_{\tau}=2.23$. The large $\tau$ spread around at the
theoretical value $\tau=1.72$ can be addressed to finite-size of
the sampling }
\end{figure}

\begin{figure}
\includegraphics[height=6cm,width=12cm]{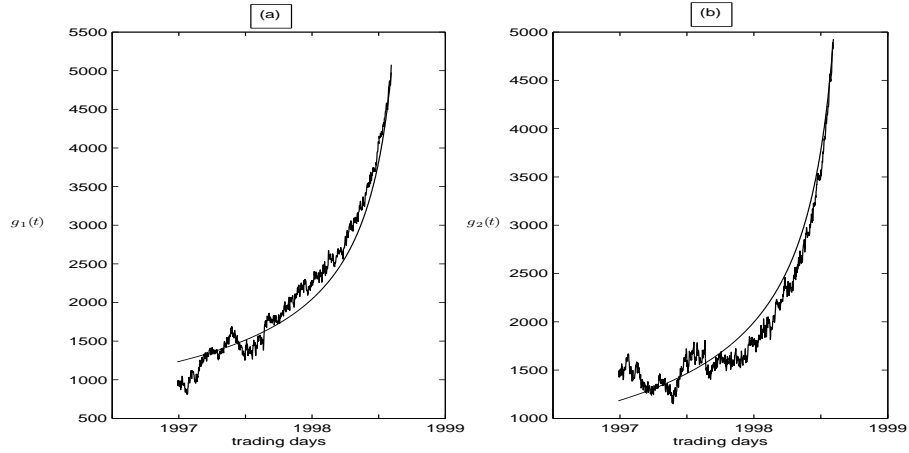}
\caption{ Sampling drawn accordingly to $g_1(t)$ (\ref{rec1})
(fig. (a)) and $g_2(t)$ (\ref{rec2}) (fig. (b)). The curves
correspond to the critical level for $\bar{f}(t)$, i.e. $g^{+}$
(fig. a) and $g^{-}$ (fig. b). $g_1(t)$ and $g_2(t)$ replicate the
deterministic exponential trend, the avalanche exponent $\tau$ and
the $H$ self-similarity exponent of the NASDAQ index (see Fig. 2
for more details on $H$ statistics)}
\end{figure}

\begin{figure}
\includegraphics[height=6cm]{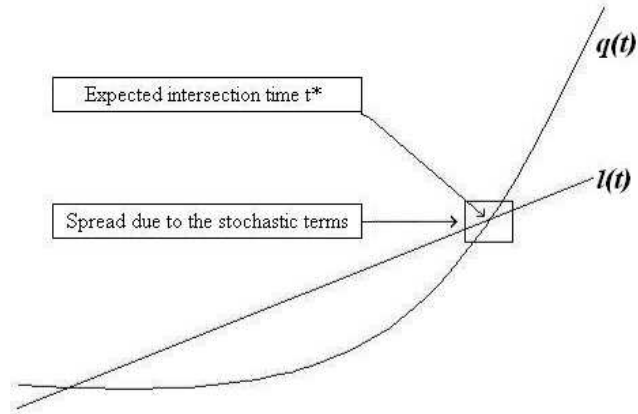}
\caption{ Line crossing }
\end{figure}

\begin{figure}
\hspace{-1cm}\includegraphics[height=8cm,width=8cm]{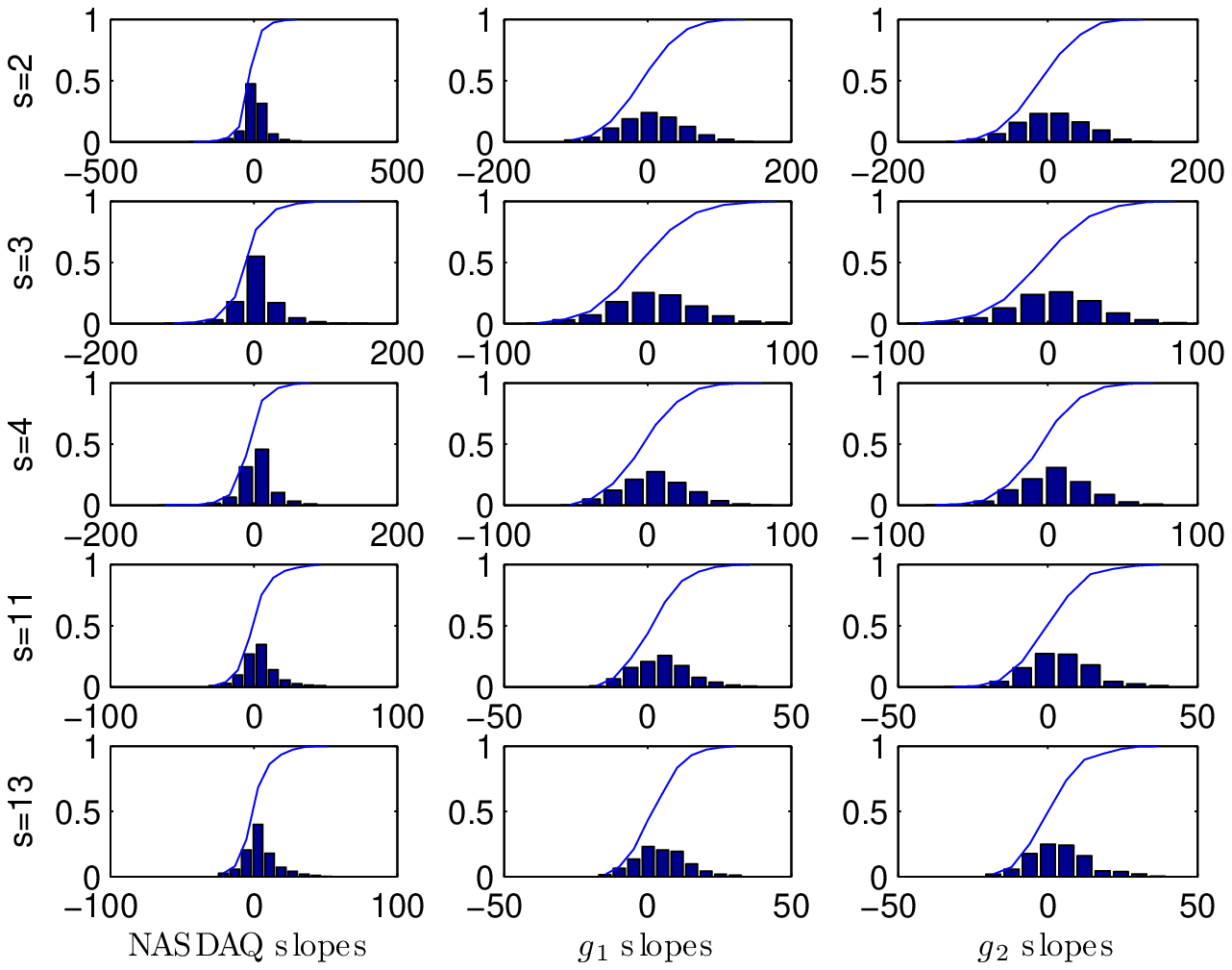}
\includegraphics[height=8cm,width=8cm]{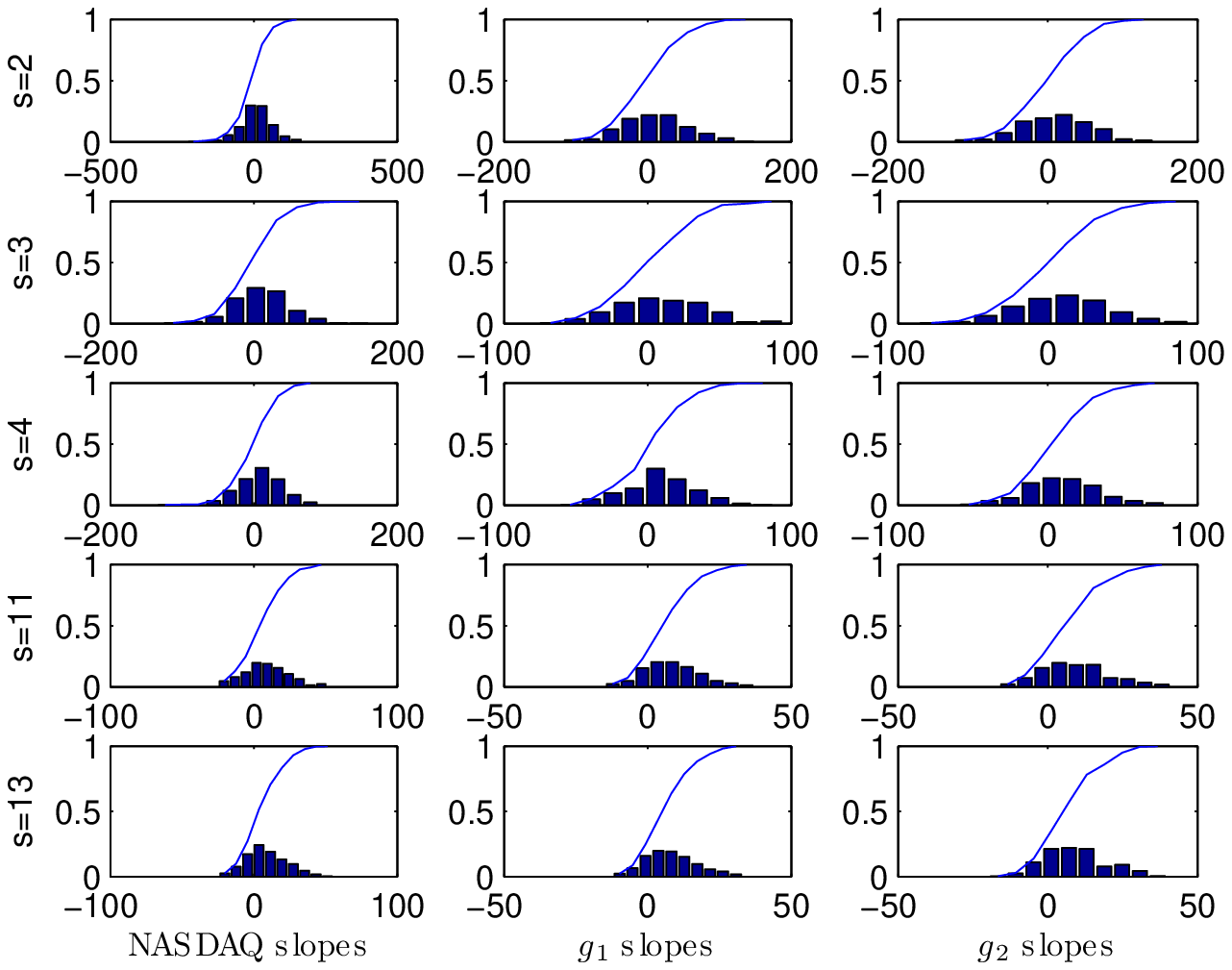}
(a)  \hspace{7cm} (b) \caption{Histograms and cumulative function
of the distribution of the slopes of lines joining $(t,
\exp(y(t)))$ and $(t+s, \exp(y(t+s)))$, $(t, g_1(t))$ and $(t+s,
g_1(t+s))$, and $(t, g_2(t))$ and $(t+s, g_2(t+s))$, respectively,
for $s=2, 3, 4, 11, 13$. (a) Analysis on the entire time series.
(b) Analysis on the time series since December 23$^{th}, 1998$}
\end{figure}

\begin{figure}
\hspace{-1cm}\includegraphics[height=8cm,width=8cm]{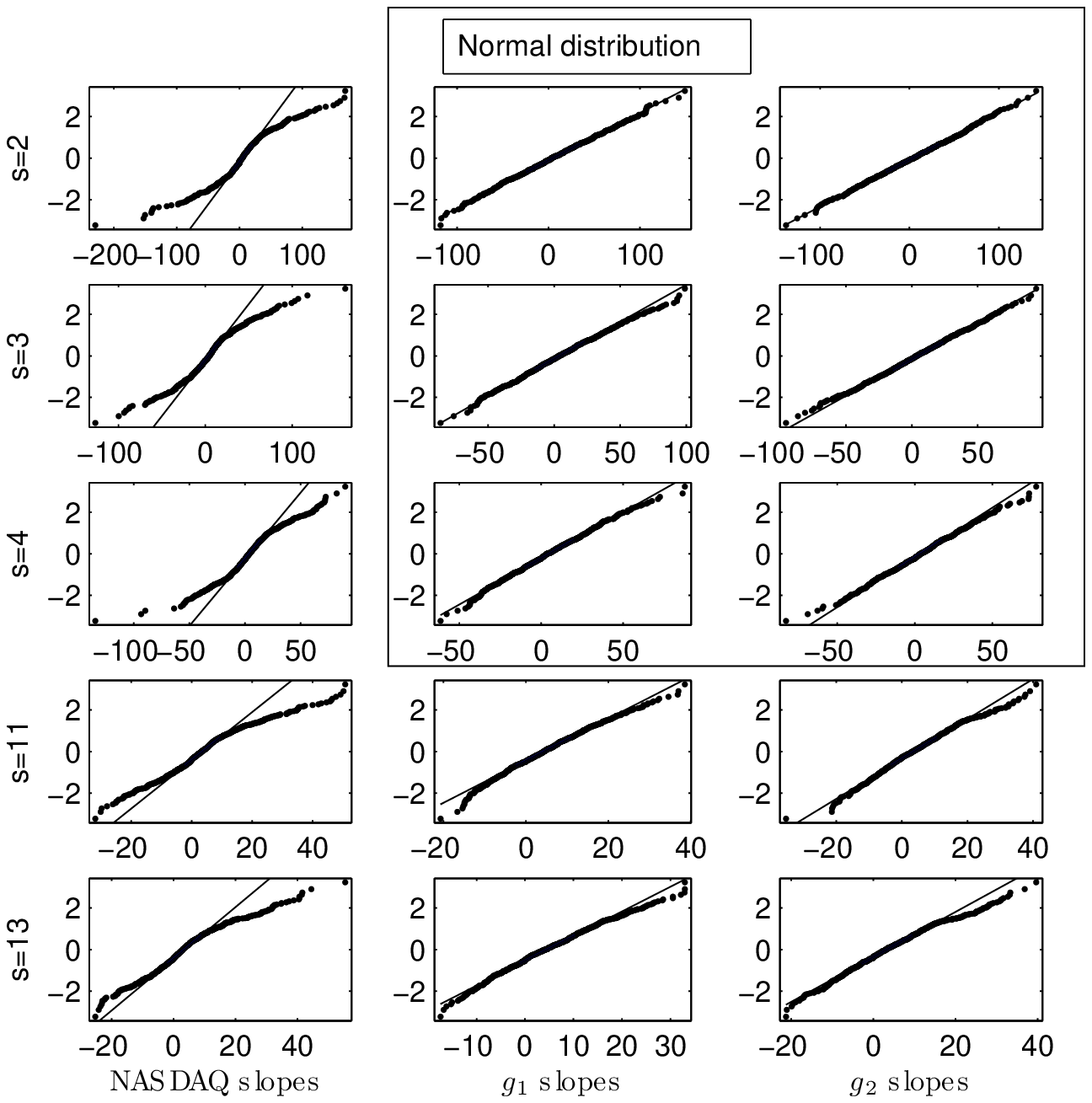}
\includegraphics[height=8cm,width=8cm]{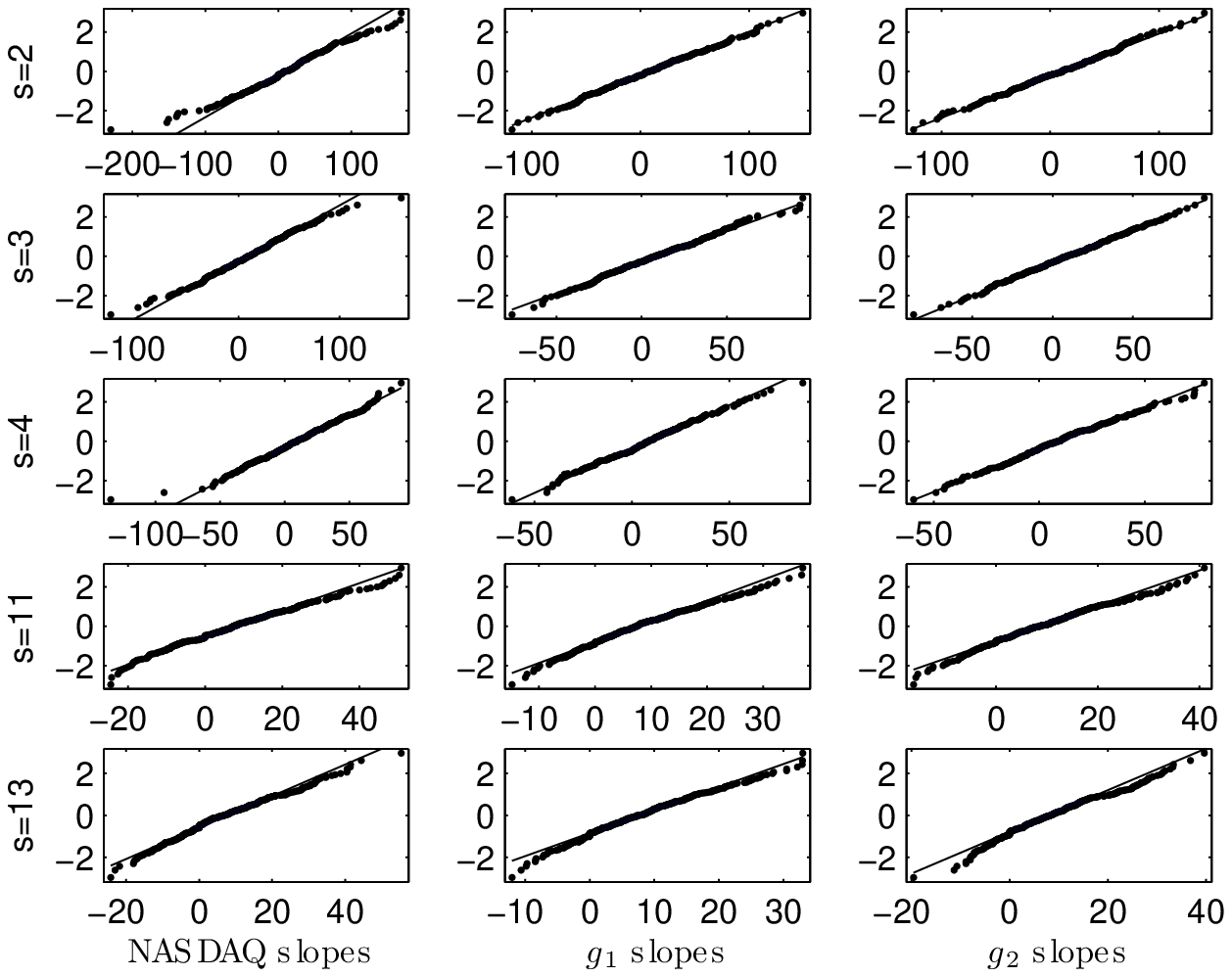}
(a)  \hspace{7cm} (b)
\caption{Normal distribution hypothesis testing of data reported
in Fig. 5. The solid line corresponds to the normal distributed
value with the same mean and variance of raw data (represented by
dots). (a) Analysis on the entire time series. Both the Lilliefors
and the Jarque-Bera test reject the normal hypothesis distribution
at $95\%$ confidence level in all the NASDAQ cases and for the
biggest values of $s=11$ and $s=13$ also on $g_1$ and $g_2$ . (b)
Analysis on the time series since December 23$^{th}$, 1998. The
Lilliefors test rejects the normal hypothesis distribution at
$95\%$ confidence level in  the NASDAQ cases $s=2$, $s=13$, and on
$g_1$ for $s=11$; the Jarque-Bera test provides the same results
given on the entire series, apart on the NASDAQ case for $s=11$,
in which it does not reject the normal hypothesis distribution}
\end{figure}

\begin{figure}
\vspace{-1cm}
(a)  \includegraphics[height=8cm]{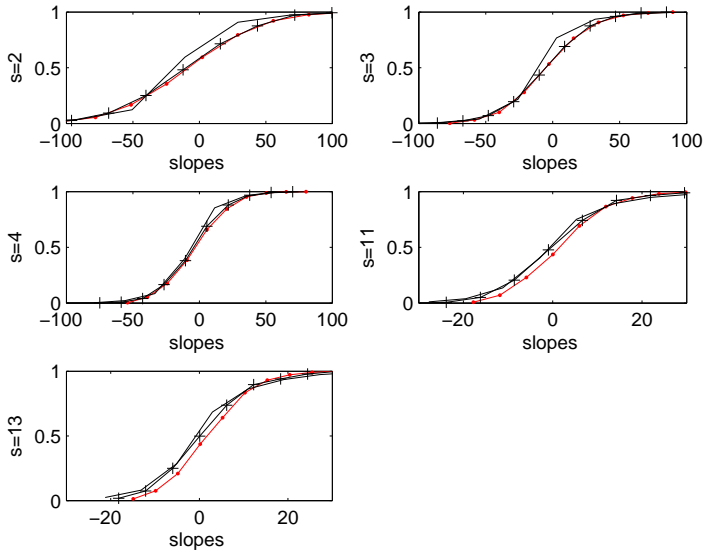}

(b)\includegraphics[height=8cm]{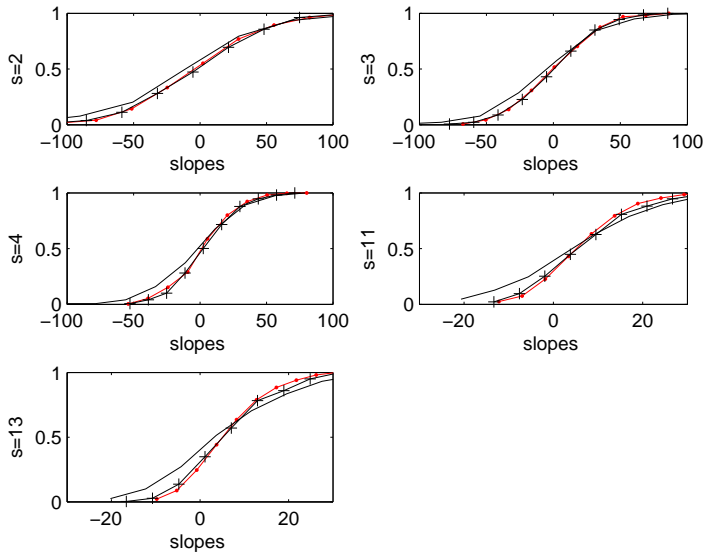}
 \caption{ For each
avalanche size $s$ the cumulative functions of the distribution of
the slopes of lines joining $(t, \exp(y(t)))$ and $(t+s,
\exp(y(t+s)))$ (solid line),  $(t, g_1(t))$ and $(t+s, g_1(t+s))$
(line with crosses), and $(t, g_2(t))$ and $(t+s, g_2(t+s))$ (line
with dots) are plotted together for   comparison. The frequency of
the slopes of $g_1$ and $g_2$ around   the mean are   lower bounds
for the NASDAQ slopes. (a) Analysis on the entire time series. The
deviation of the NASDAQ from the normal distribution and the
leptokurtosis  is evidenced at most for $s=2$.   (b) The same
analysis performed on the time series since December 23$^{th}$,
1998. Deviation of the NASDAQ from the normal distribution  is
best evidenced
 for $s=2$}
\end{figure}

\begin{figure}
\includegraphics[height=6cm]{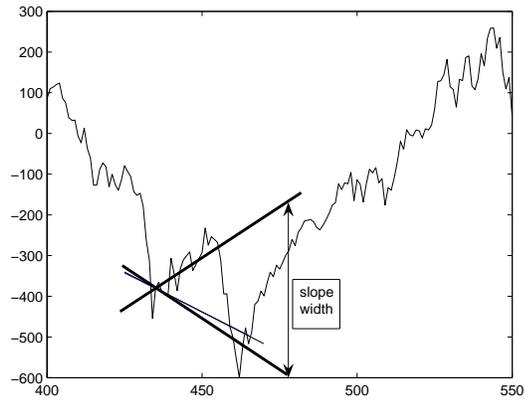}
\caption{ An example of 10\% interval of slopes (upper and lower
lines) together with a line actually joining $(t, g_1(t))$ and
$(t+s, g_1(t+s))$ with $s=2$}
\end{figure}

\begin{figure}
\includegraphics[height=6cm]{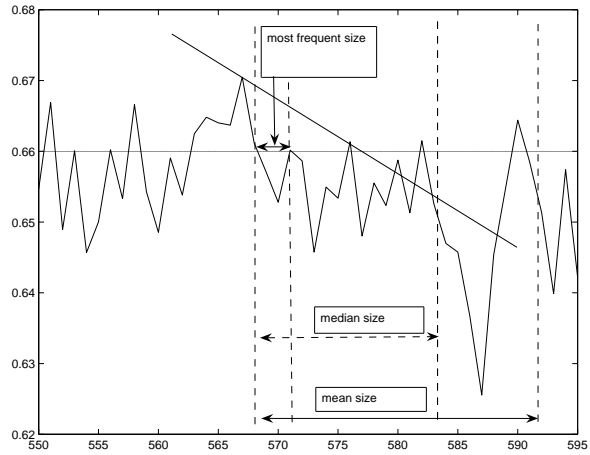}
\caption{ A sampling of $\bar{f}^d(t)$. The most frequent size of
an avalanche is smaller than the median size, that is smaller than
the mean }
\end{figure}

\end{document}